\documentclass[12pt]{article}
\usepackage{graphicx}
\usepackage[breaklinks]{hyperref}
\usepackage{cite}
\usepackage[utf8]{inputenc}
\pdfoutput=1

\newcommand\pubnumber{}
\newcommand\pubdate{}

\def\infn{INFN, Sezione di Roma\\
P.le A. Moro, 2, I-00185 Rome, Italy}
\def\support{\footnote{The research leading to these results
has received funding from the European Research Council under the European 
Union's Seventh Framework Programme (FP/2007-2013) / ERC Grant
Agreements n. 279972 ``NPFlavour'' and n. 267985 ``DaMeSyFla''.}}

\def\Title#1{\begin{center} {\Large #1 } \end{center}}
\def\Author#1{\begin{center}{ \sc #1} \end{center}}
\def\Address#1{\begin{center}{ \it #1} \end{center}}

\newcommand\pubblock{\rightline{\begin{tabular}{l} \pubnumber\\
         \pubdate  \end{tabular}}}
\newenvironment{Abstract}{\begin{quotation}  }{\end{quotation}}
\newenvironment{Presented}{\begin{quotation} \begin{center} 
             PRESENTED AT\end{center}\bigskip 
      \begin{center}\begin{large}}{\end{large}\end{center} \end{quotation}}
\def\Acknowledgements{\bigskip  \bigskip \begin{center} \begin{large}
             \bf ACKNOWLEDGEMENTS \end{large}\end{center}}

\def\beq{\begin{equation}}
\def\eeq#1{\label{#1}\end{equation}}
\def\eeqn{\end{equation}}

\def\beqa{\begin{eqnarray}}
\def\eeqa#1{\label{#1}\end{eqnarray}}
\def\eeqan{\end{eqnarray}}

\let\bar=\overbar

\def\Dslash{\not{\hbox{\kern-4pt $D$}}}
\def\dslash{\not{\hbox{\kern-2pt $\del$}}}

\def\msb{{\bar{\ssstyle M \kern -1pt S}}}

\begin{document}
\begin{titlepage}
\pubblock

\vfill
\Title{CHARM-2015 Theory Summary}
\vfill
\Author{Luca Silvestrini\support}
\Address{\infn}
\vfill
\begin{Abstract}
I present a brief theory overview of the CHARM-2015 conference.
\end{Abstract}
\vfill
\begin{Presented}
The 7th International Workshop on Charm Physics (CHARM 2015)\\
Detroit, MI, 18-22 May, 2015
\end{Presented}
\vfill
\end{titlepage}
\def\thefootnote{\fnsymbol{footnote}}
\setcounter{footnote}{0}
%

\section{Introduction}

CHARM-2015 has been a most lively conference, with more than twenty
very interesting theoretical presentations
\cite{piccinini,lebed,esposito,prelovsek,mohler,molina,padmanath,bodwin,hzhang,luchinsky,qiu,vogt,yu,zhao,nahrgang,lytle,mateu,petriello,zzhang,gamiz,kagan,fajfer,deboer,zupan,paul,schacht,kosnik,lai}. Due
to my limitations, as well as for reasons of space, I cannot possibly
do justice to all of them, so rather than a full-fledged summary I
will only give my personal view of the conference, referring the
interested reader to the contributions collected in this volume.

\begin{figure}[t]
\centering
\includegraphics[width=0.8\textwidth]{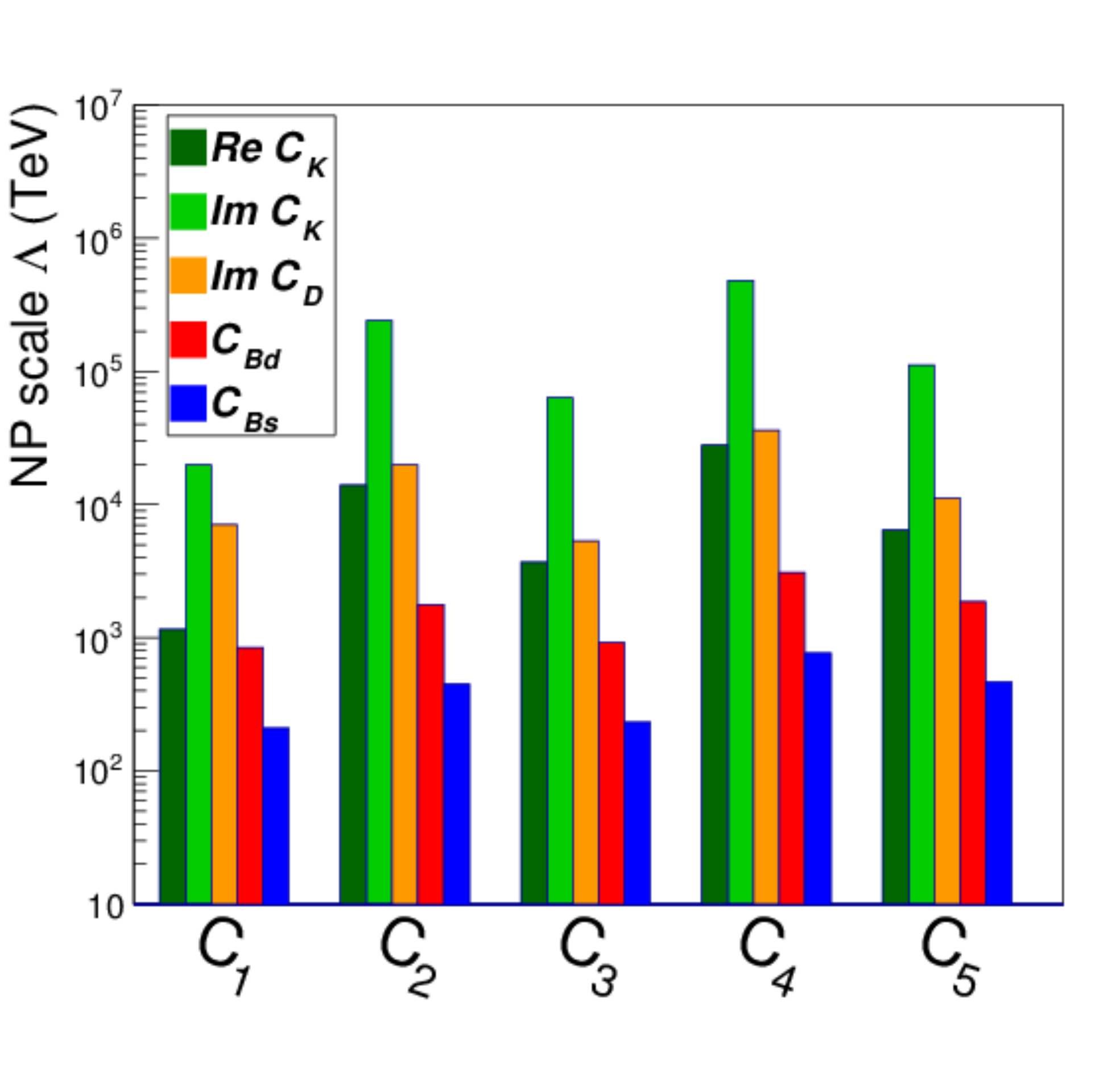} 
\caption{Bounds on the NP scale $\Lambda$ from $\Delta F=2$
  decays. See ref.~\cite{Bona:2007vi} for details.}
\label{fig:df2}
\end{figure}

The spectacular experimental progress that we have witnessed in the
past few years is leading us in the precision charm physics era,
calling for substantial theoretical advances to fully exploit the
wealth of available data. Charm physics is now at the forefront of New
Physics (NP) searches, allowing us to probe energies as high as $10^4$
TeV's \cite{Carrasco:2014uya} (see Fig.~\ref{fig:df2}), with ample room
for sizable improvements, both from the theoretical and experimental
point of view.

In Sec.~\ref{sec:properties} I quickly report on recent progress in
the determination of charm properties: spectroscopy, production, mass,
decay constants and form factors. In Sec.~\ref{sec:clean} I discuss
NP-sensitive, theoretically clean observables, such as CP violation in
$D$ mixing and a few rare decays. Finally, in Sec.~\ref{sec:dirty} I
mention several potentially NP-sensitive but theoretically challenging
processes, such as CP violation in nonleptonic $D$ decays and more
rare decays.

\section{Charm properties}
\label{sec:properties}

\subsection{Spectroscopy}
\label{sec:spectro}

Twelve years after the $X(3872)$ discovery, the $c \bar{c}$ spectrum
has been widely explored: all the states below the open charm
threshold have been identified, all the $1^{--}$ states are filled,
but the long-standing problem of understanding the structure of
exotica is still open \cite{piccinini}. While considering exotica as
loosely bound charmed meson molecules gives an economic description of
several exotic states very close to threshold, this explanation is
challenged by prompt production at the LHC. On the other hand, the
description of exotica in terms of compact tetraquarks implies the
prediction of (too) many additional states, depending on the details
of the diquark interaction (a very interesting subject \textit{per se}
\cite{lebed}), and it is supported by the observation of new charged
states. Studying decays in specific channels could discriminate
between models \cite{esposito}, and more experimental data will
certainly help in finally clarifying this open issue. 

The spectroscopy of $c \bar{c}$ states can also be studied on the
lattice \cite{prelovsek}. While precision results in excellent
agreement with experiment have been obtained for the states well below
threshold, the situation becomes problematic when the energy raises
above threshold. Correlation functions in the Euclidean are always
dominated by the state with the lowest energy, preventing the study of
interacting multi-meson states above threshold
\cite{Maiani:1990ca}. Finite volume effects allow to overcome this
limitation for two-meson states \cite{Lellouch:2000pv}, but
substantial progress is still needed for three-meson states. In spite
of these difficulties, first studies of exotic $X$, $Y$ and $Z$ states
have been carried out
\cite{Lang:2015sba,Prelovsek:2013cra,Lee:2014uta,mohler,molina}, 
although no firm conclusion on the nature of these states has been
reached yet \cite{Padmanath:2015era,Guerrieri:2014nxa}. Lattice
studies of charmed baryon spectroscopy are in a similar situation: for
ground states there is good agreement between lattice results and
experiments, while the study of excited states is really challenging
\cite{padmanath}. 

\subsection{Charm production}
\label{sec:prod}

Let us now briefly review recent progress on charm production,
starting with quarkonium production in the vacuum
\cite{bodwin,hzhang,luchinsky}.  The cross-section for quarkonium
production at high-$p_T$ is expected to factorize, order by order in
an expansion in the velocity $v$, into the product of the
short-distance partonic cross section, convoluted with the pdf, times
the long-distance probability for a $Q \bar{Q}$ pair to evolve into a
quarkonium state \cite{Bodwin:1994jh}. Nonrelativistic-QCD
Factorization has been proven up to NLO, but an all-orders proof is
still missing. Predictions depend on the long-distance matrix
elements, which are assumed to be universal and must be determined
phenomenologically. Combining factorization with fragmentation at
leading $(1/p_T^4)$ \cite{Collins:1981uw} and next-to-leading
$(m_Q^2/p_T^6)$ \cite{Kang:2011zza} power, one can estimate the
dominant effects at large $p_T$, obtaining a good description of
$J/\psi$ hadroproduction. However, $J/\psi$ photoproduction at Hera
and $\eta_c$ hadroproduction at LHCb still appear problematic
\cite{bodwin}. Quarkonium production in matter is even more
problematic and remains an open problem, although considerable effort
is being made to clarify this subject \cite{qiu,vogt,yu,zhao}. Open
charm production in matter is presently well described by a
considerable number of models, and more observables are needed to
discriminate between them \cite{nahrgang}.

\subsection{Charm quark mass and Yukawa coupling}
\label{sec:mass}

\begin{figure}[t]
\centering
\includegraphics[width=0.53\textwidth]{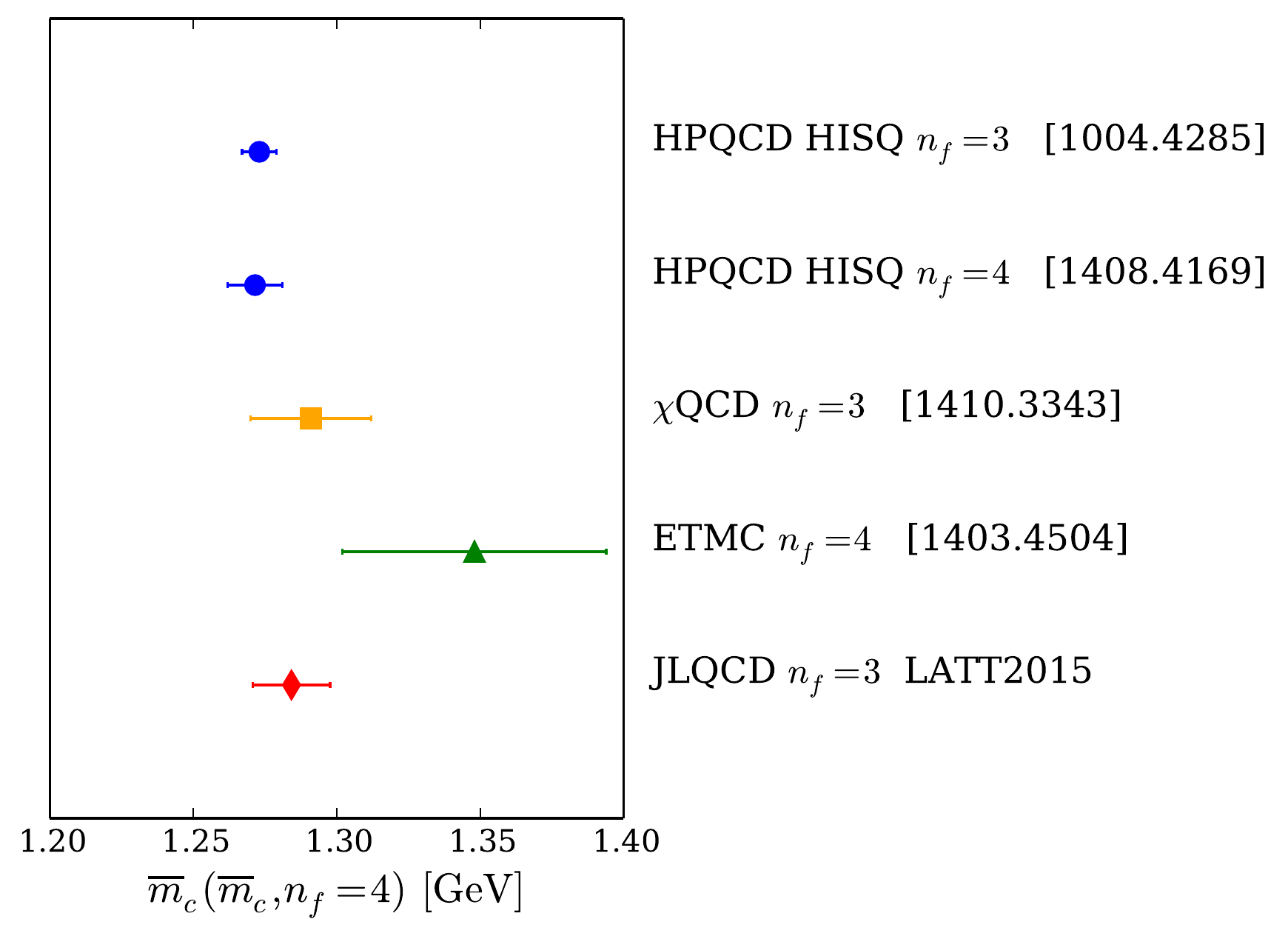}
\includegraphics[width=0.4\textwidth]{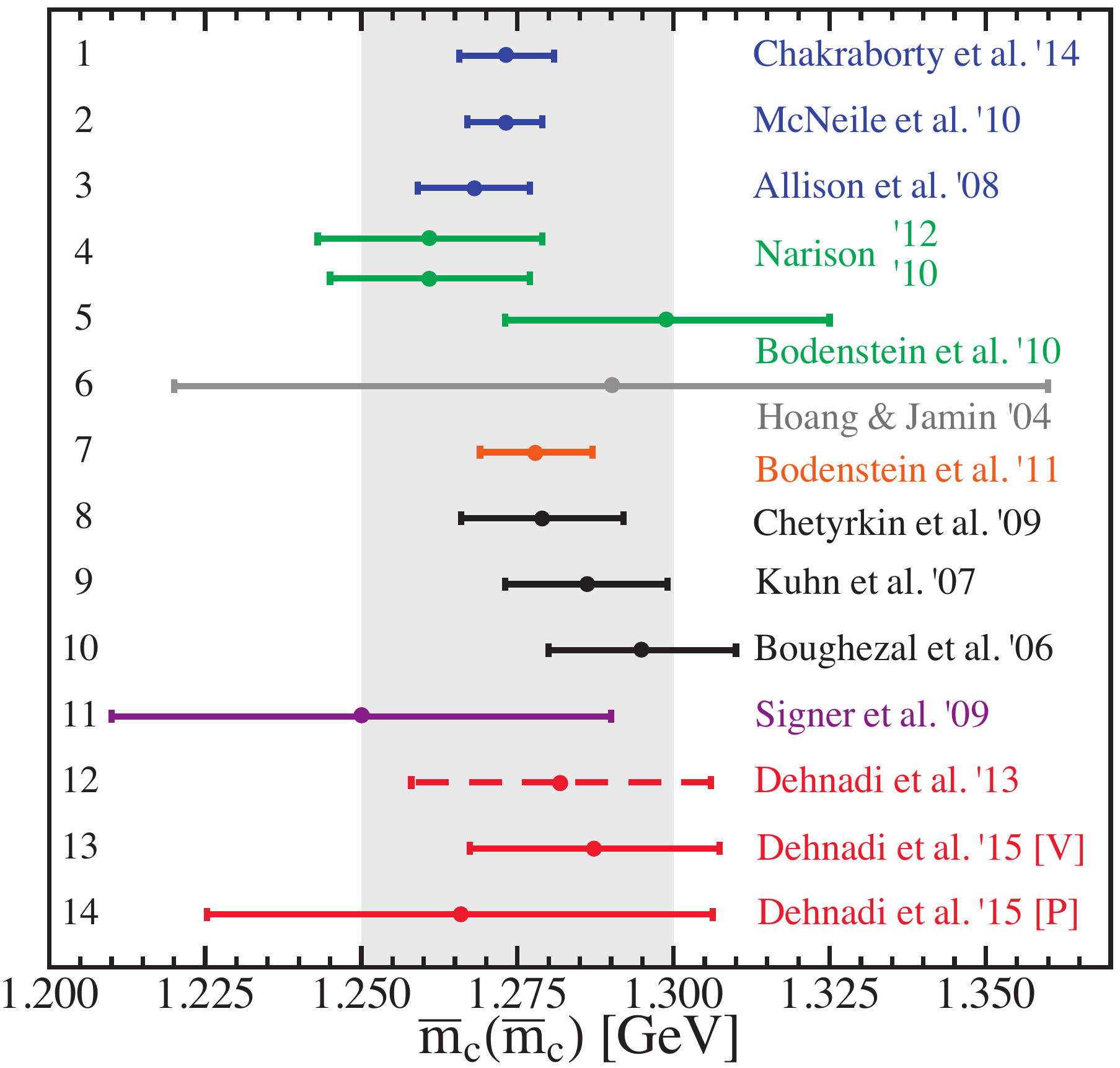}
\caption{Comparison of recent determinations of the charm quark mass
  from Lattice QCD (left panel, from ref.~\cite{lytle}) and from QCD
  sum rules (right panel, from ref.~\cite{mateu}).}
\label{fig:mc}
\end{figure}

The charm quark mass can be determined with nonperturbative methods
such as Lattice QCD or QCD sum rules. Remarkable progress has been
recently achieved in both approaches. Lattice QCD calculations with
three or four active flavours have been performed by several
collaborations, using different actions, renormalization procedures
and methods; QCD sum rules computations include terms of
$\mathcal{O}(\alpha_s^3)$ and several tests can be performed on the
convergence of the perturbative expansion. A collection of recent
results obtained in both approaches is presented in Fig.~\ref{fig:mc};
more details can be found in refs.~\cite{lytle,mateu}.

The determination of quark Yukawa couplings and their relation with
quark masses is a crucial test of the validity of the Standard
Model. In this respect, a direct determination of the charm Yukawa
coupling would be extremely important. Unfortunately, this is a
formidable task, requiring very high integrated luminosity
\cite{petriello,zzhang}. A promising approach to the determination of
the charm Yukawa coupling is via $h \to J/\psi \gamma$, using the
interference of direct and indirect production, which is theoretically
clean and could give interesting results with 3 ab$^{-1}$
\cite{Bodwin:2013gca,Bodwin:2014bpa}. 

\subsection{Charmed meson decay constants and form factors}
\label{sec:decay}

\begin{figure}[!htbp]
\centering
\includegraphics[width=0.8\textwidth]{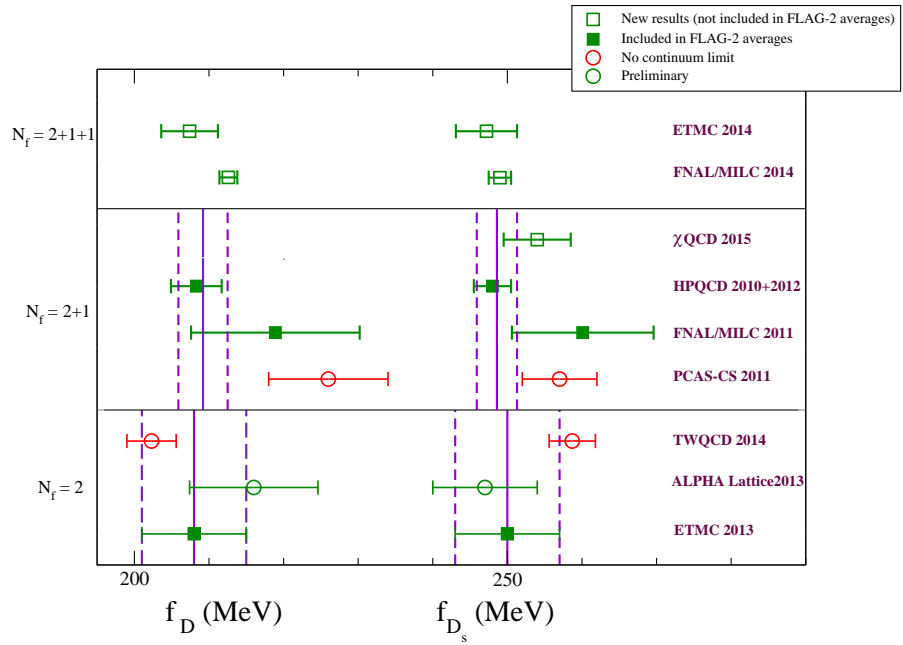} 
\includegraphics[width=0.8\textwidth]{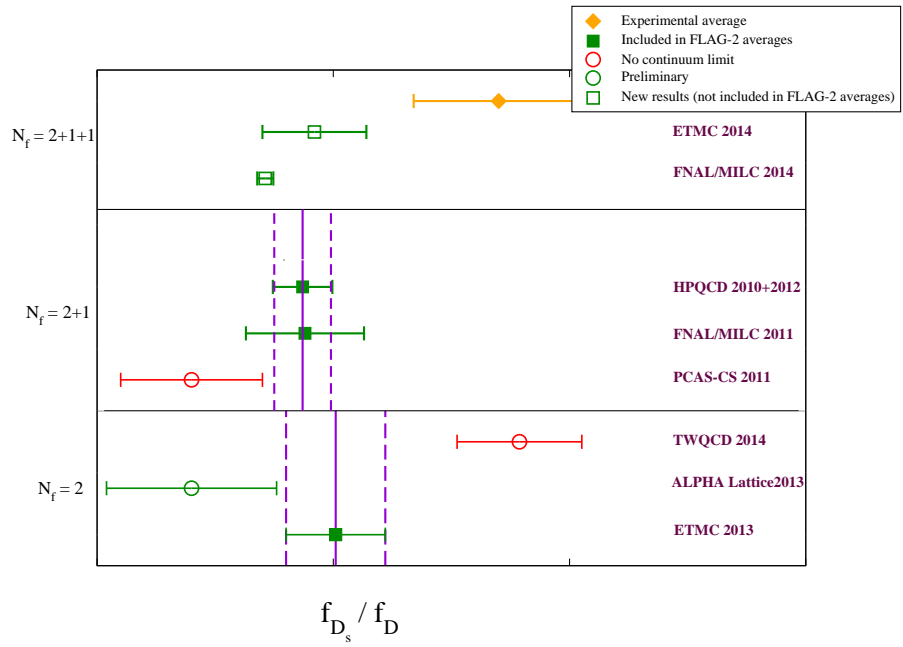}
\caption{Recent determinations of the $D$ and $D_s$ decay constants
  from lattice QCD \cite{gamiz}.}
\label{fig:fdfds}
\end{figure}
\begin{figure}[!htbp]
\centering
\includegraphics[width=0.6\textwidth]{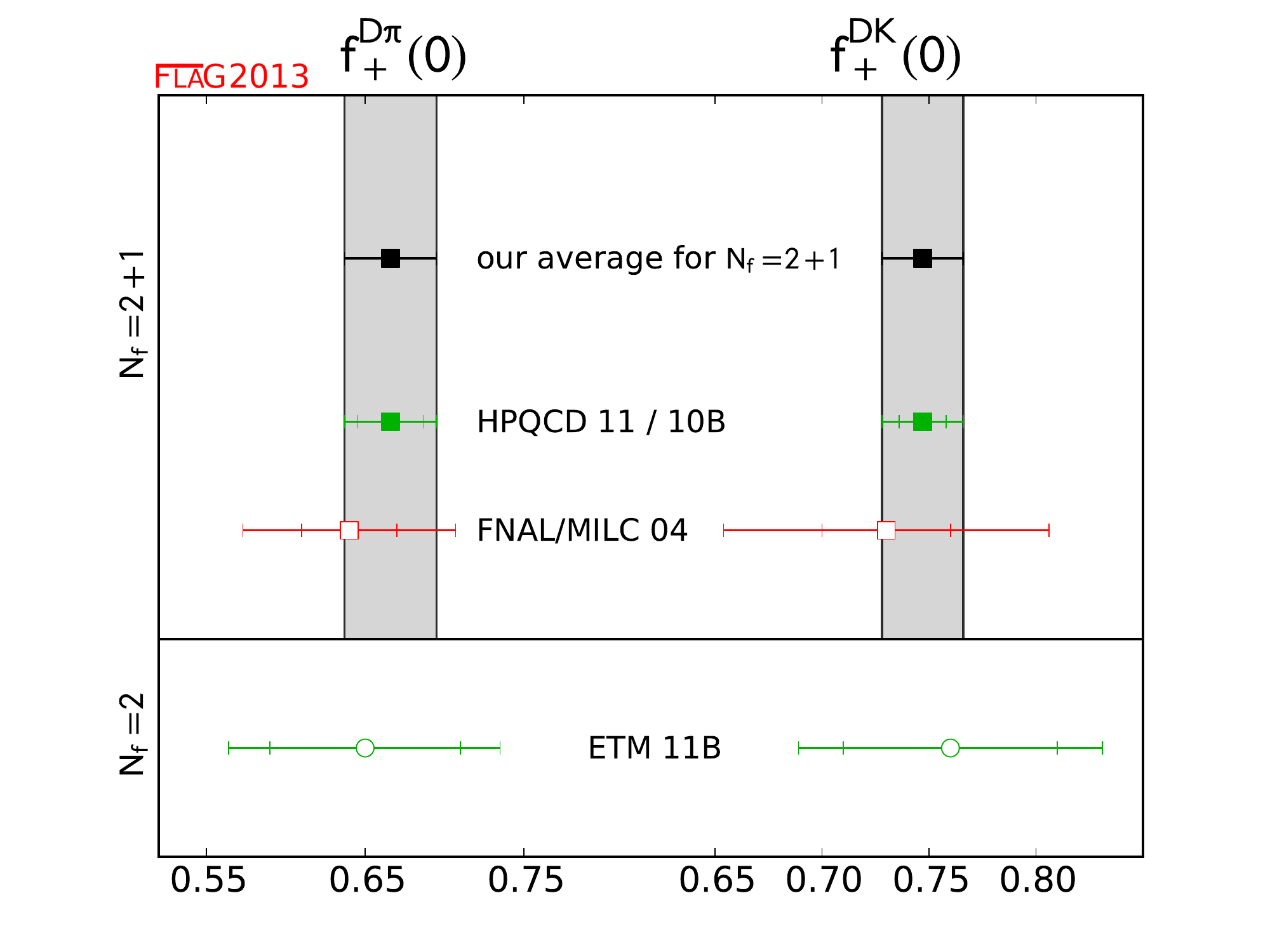}
\includegraphics[width=0.45\textwidth]{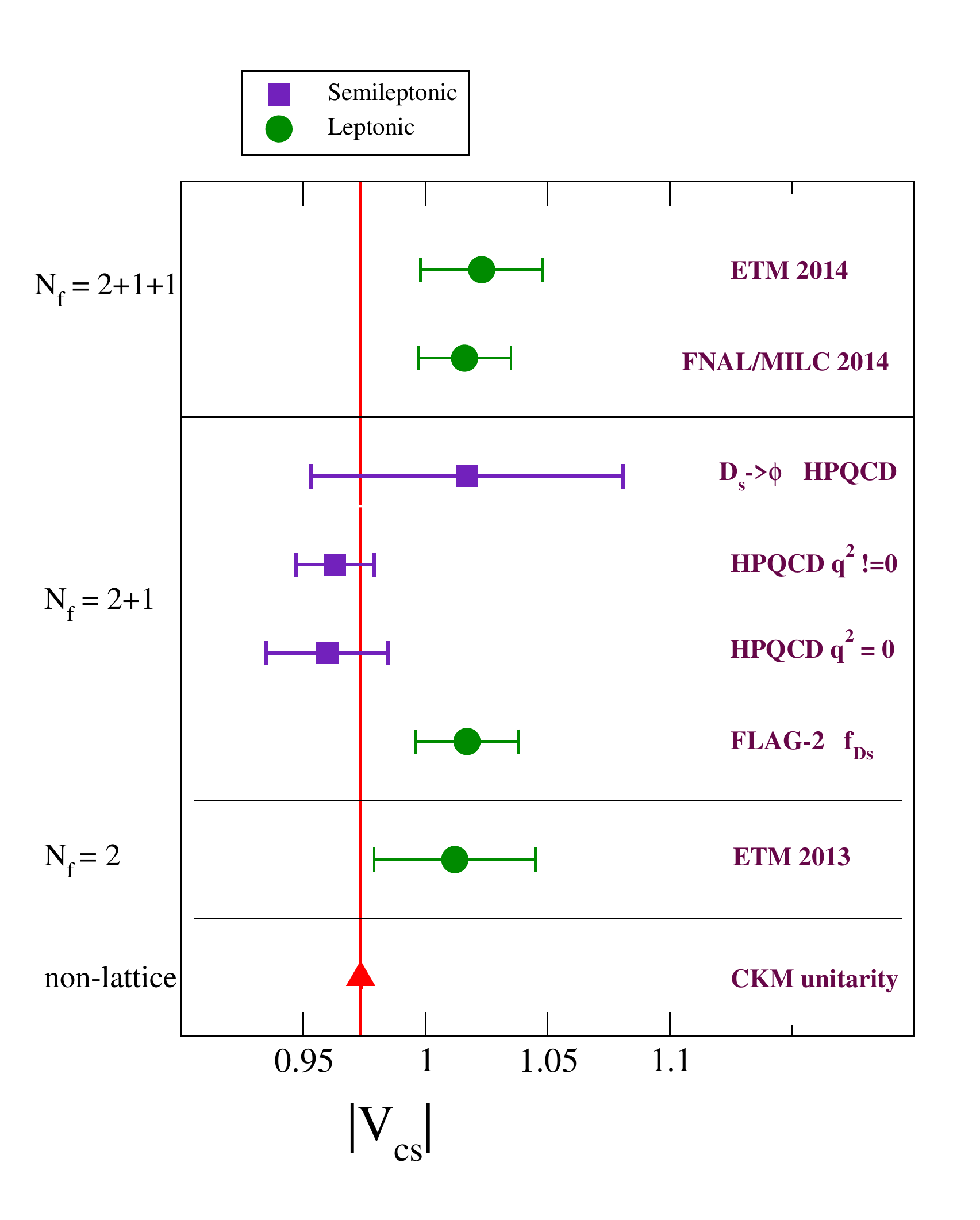}
\includegraphics[width=0.45\textwidth]{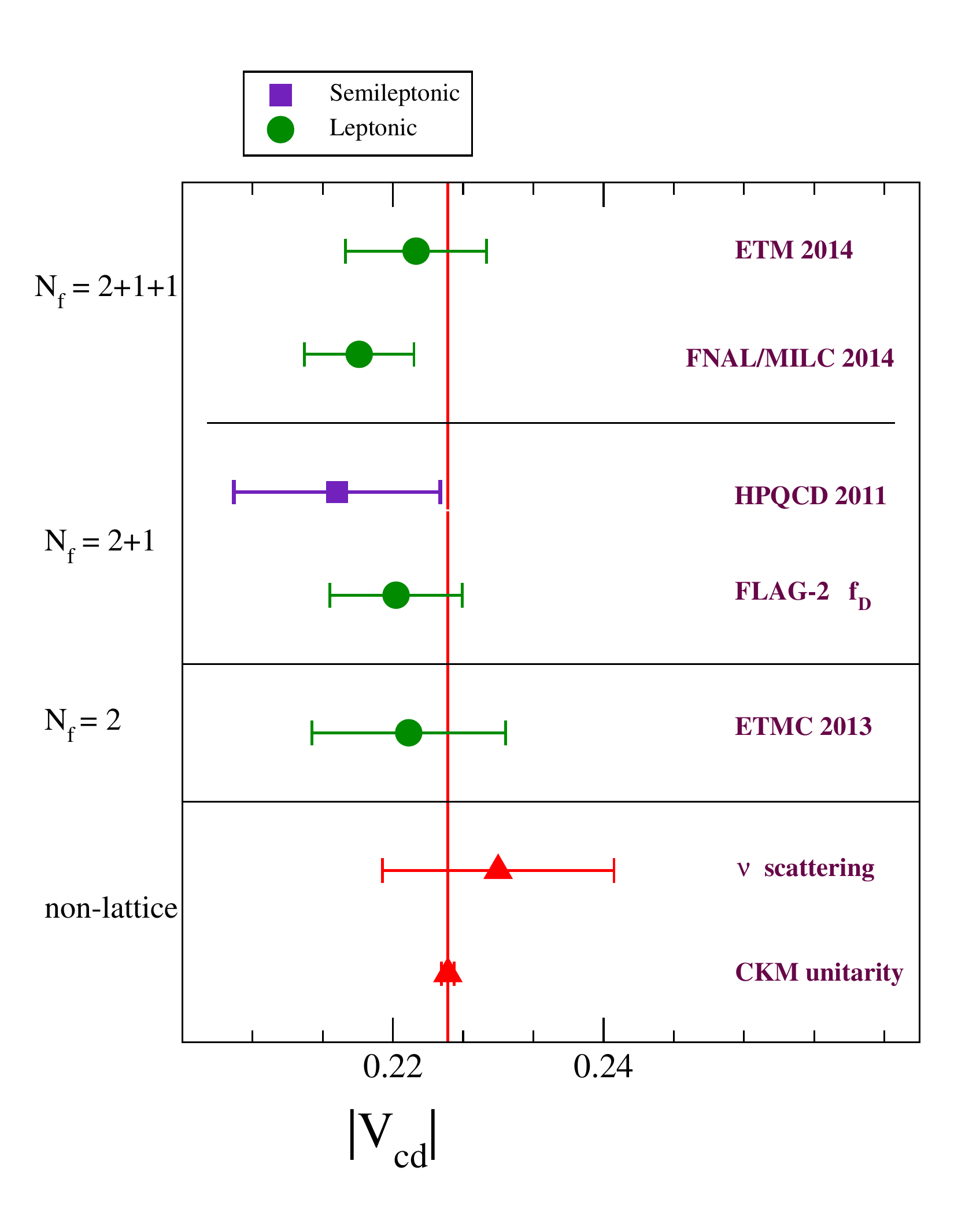}
\caption{Recent determinations of the semileptonic form factors (top) and of CKM
  unitarity (bottom) from lattice QCD
  \cite{gamiz,Aoki:2013ldr}.}
\label{fig:semil}
\end{figure}

Considerable progress is also taking place in the precision
determination of charmed mesons decay constants and form factors in
Lattice QCD. Figures \ref{fig:fdfds} and \ref{fig:semil} summarizes the
current averages from the Flavour Lattice Averaging Group (FLAG)
\cite{Aoki:2013ldr} as well as more recent calculations not yet
included in the FLAG averages \cite{gamiz}. The experimental numbers
for the decay constants are $f_{D_s} = 257.5 \pm 4.6$ MeV and
$f_{D_s}/f_D = 1.258 \pm 0.038$ \cite{pdg}. The $n_f = 2$ FLAG
averages and the recent ETMC $n_f = 2 + 1 + 1$ results
\cite{Dimopoulos:2013qfa} are in fair agreement with data, while some
tension is seen comparing the $n_f = 2 + 1$ FLAG averages and the
recent FNAL/MILC $n_f = 2 + 1 + 1$ results \cite{Bazavov:2014lja}.

\section{NP-sensitive, theoretically clean processes}
\label{sec:clean}

CP-violation in $\Delta F=2$ processes is the most sensitive probe of
NP, reaching NP scales as high as $\mathcal{O}(10^5)$ TeV for generic
flavour structures and $\mathcal{O}(1)$ couplings (see
Fig.~\ref{fig:df2}). Thanks to the recent experimental and theoretical
improvements, CP violation in $D$ mixing is giving the second best
constraint on NP; furthermore, combining bounds from $K$ and $D$
mixing allows to constrain several NP models much more effectively
than considering bounds from individual processes. 

From the theoretical point of view, $D$ mixing is described in terms
of the dispersive and absorptive mixing amplitudes $M_{12}$ and
$\Gamma_{12}$. In the SM, both amplitudes are dominated by long
distance contributions and thus not calculable at present
\cite{yu}. NP contributions to $\Gamma_{12}$ are expected to be
negligible, while NP could give large short-distance contributions to
$M_{12}$, which can be accurately computed using matrix elements
computed on the lattice \cite{Carrasco:2014uya}. The observables
related to the mixing amplitude are $\vert M_{12} \vert$, $\vert
\Gamma_{12} \vert$ and $\Phi_{12} = \arg (\Gamma_{12}/M_{12})$. Being
Flavour Changing Neutral Current (FCNC) processes, the mixing
amplitudes are GIM suppressed, due to the unitarity of the CKM matrix,
and in particular to the unitarity relation $\lambda_d + \lambda_s +
\lambda_b = 0$, where $\lambda_{d_i} = V_{cd_i}
V^\star_{ud_i}$. Unitarity allows to eliminate $\lambda_d$;
furthermore, we can choose $\lambda_s$ to be real, so that all CPV is
generated by terms proportional to $\lambda_b$ and thus suppressed by
$r = \mathrm{Im} \lambda_b/\lambda_s \sim 6.5\cdot 10^{-4}$. Denoting by
$f_{d_id_j}$ the loop amplitude with $d_i \bar{d}_j$ intermediate
states, both $M_{12}$ and $\Gamma_{12}$ have the following structure:
\begin{equation}
  \label{eq:gimsu3}
  \lambda_s^2 (f_{dd}+f_{ss}-2 f_{ds}) + 2 \lambda_s \lambda_b (f_{dd}
  + f_{bs} - f_{bd} - f_{sd}) + \mathcal{O}(\lambda_b^2)\,.
\end{equation}
From eq.~(\ref{eq:gimsu3}) it is evident that for the dominant
long-distance contributions GIM suppression coincides with
$\mathrm{SU}(3)$ suppression. Indeed, $\mathrm{SU}(3)$ can serve as a
guiding principle to estimate the size of the two terms in
eq.~(\ref{eq:gimsu3}). To this aim, it is useful to rewrite them in
terms of U-spin quantum numbers:
\begin{equation}
  \label{eq:uspin}
  \lambda_s^2 (\Delta \mathrm{U} = 2)  + 2 \lambda_s \lambda_b 
  (\Delta \mathrm{U} = 1 + \Delta \mathrm{U} = 2) +
  \mathcal{O}(\lambda_b^2) \sim \lambda_s^2 \epsilon^2 + 2 \lambda_b
  \lambda_s \epsilon\,.
\end{equation}
We see that CP violating effects are expected to arise at the level of
$r/\epsilon \sim 2 \cdot 10^{-3} \sim 1/8^\circ$ for nominal
$\mathit{SU}(3)$ breaking $\epsilon = 30\%$. Given the present
experimental errors, it is therefore perfectly adequate to assume real
$M_{12}$ and $\Gamma_{12}$ in the SM as well as real decay
amplitudes, allowing to fit all $D$-mixing data using the universal
parameters \cite{Ciuchini:2007cw,Kagan:2009gb}
\begin{eqnarray}
  \label{eq:dmixpars}
  &&x\sim 2 \vert M_{12} \vert/\Gamma\,,\qquad y \sim \vert \Gamma_{12}
\vert/\Gamma \\
  && \vert q/p \vert = \frac{\sqrt{4 \vert M_{12}\vert^2 + \vert \Gamma_{12}
  \vert^2 + 2\vert M_{12}\vert\vert \Gamma_{12}\vert \sin
  \Phi_{12}}}{\Gamma \sqrt{x^2+y^2}}\,, \;
  \phi = \arg \left( y + i\, (1-\vert q/p \vert) x\right)\,.
\nonumber
\end{eqnarray}
A possible NP-induced phase in $M_{12}$ (we expect
NP to give negligible contributions to $\Gamma_{12}$) would result in
$\vert q/p \vert -1 \neq 0$ and in $\phi \neq 0$.

\begin{figure}[htb]
\centering
\includegraphics[width=0.3\textwidth]{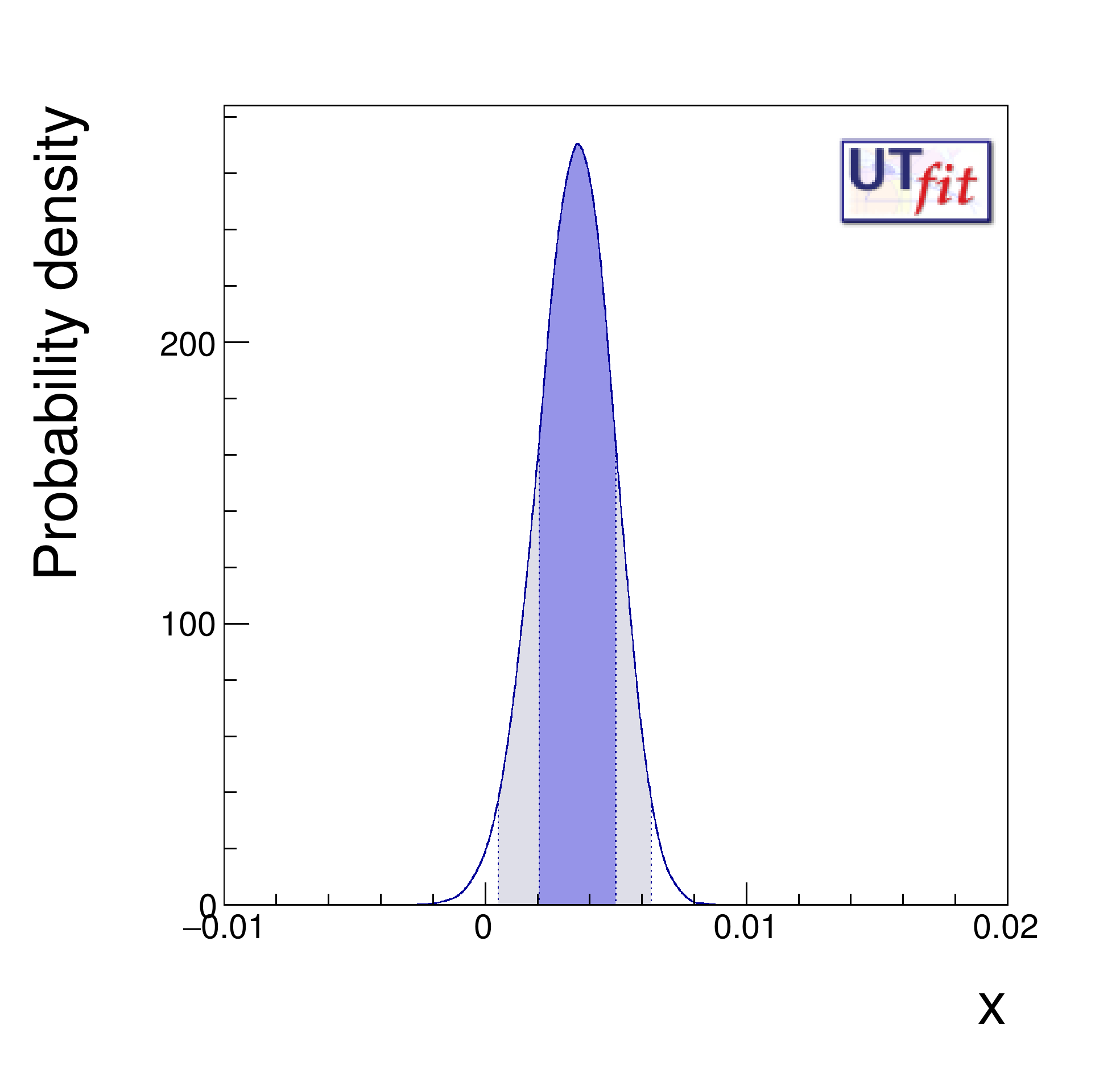} 
\includegraphics[width=0.3\textwidth]{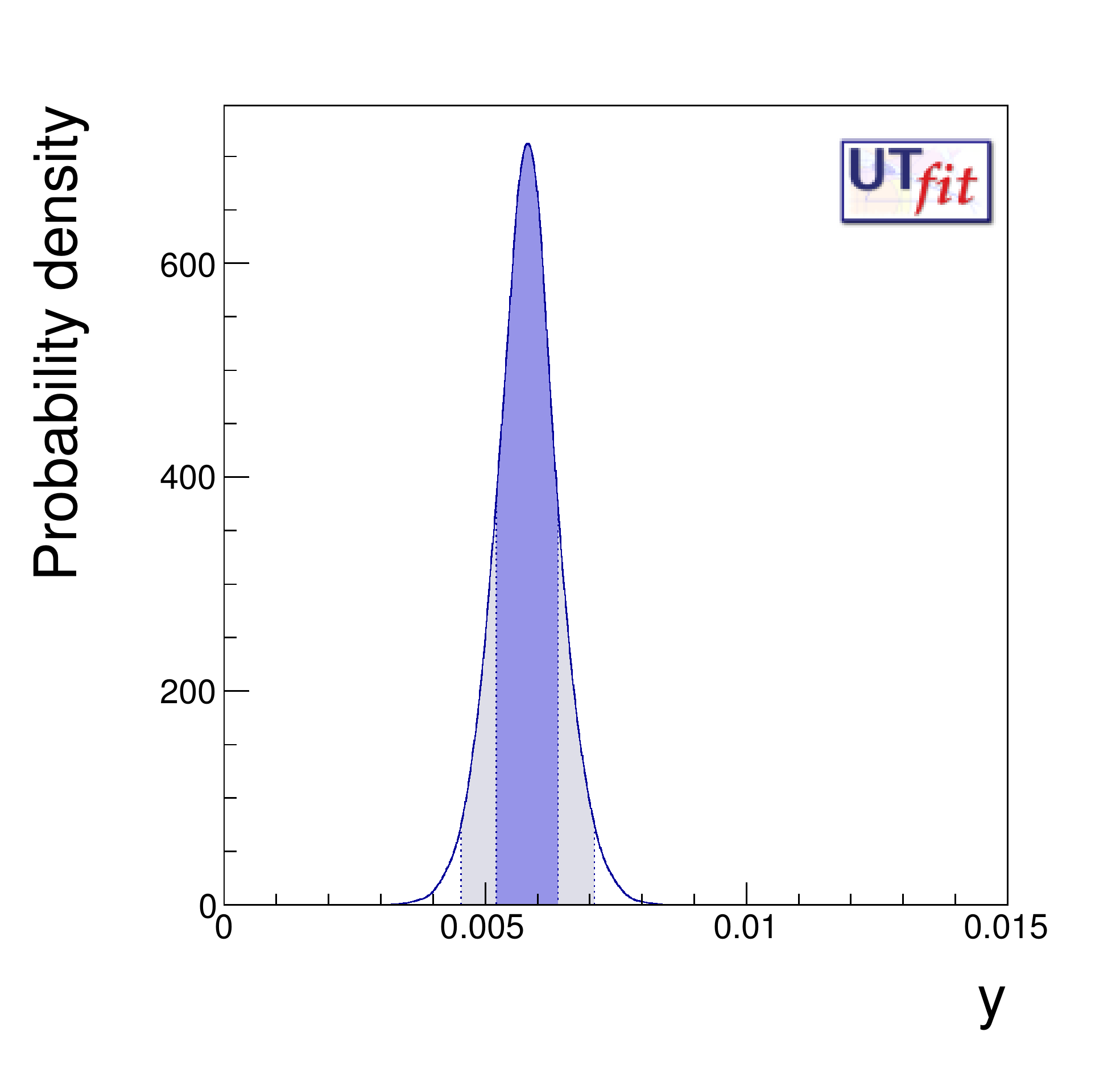} 
\includegraphics[width=0.3\textwidth]{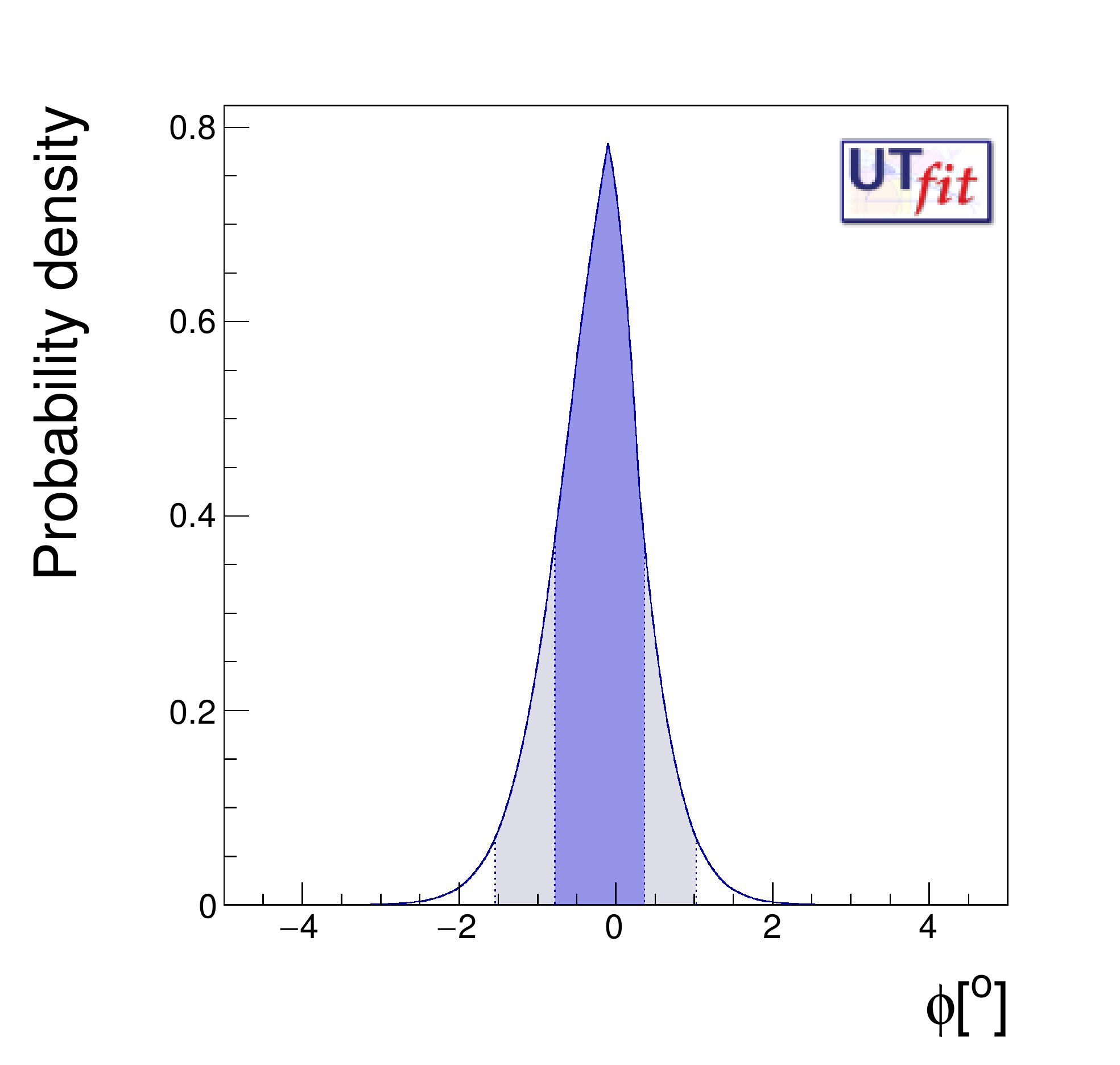} 
\includegraphics[width=0.3\textwidth]{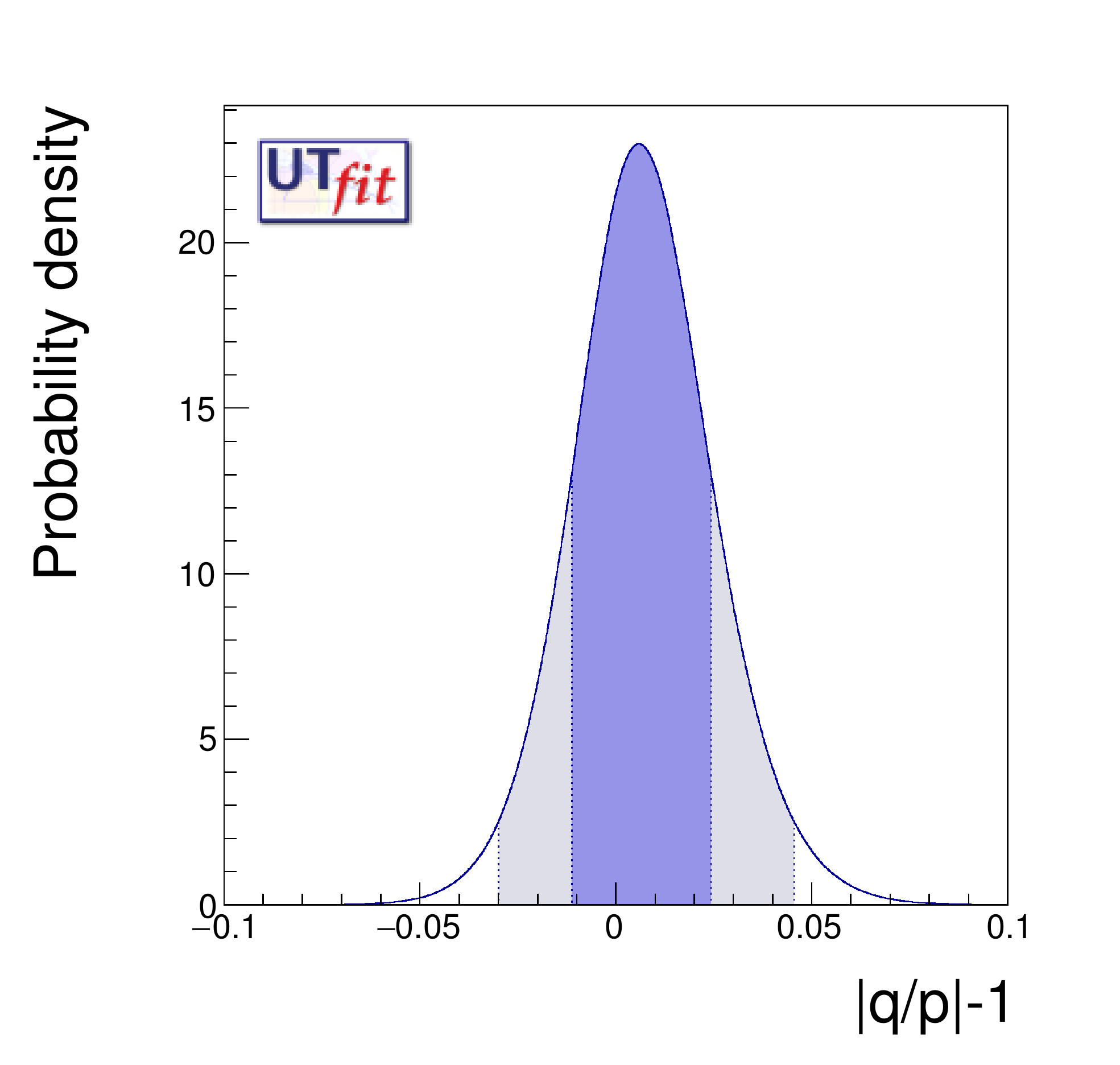} 
\includegraphics[width=0.3\textwidth]{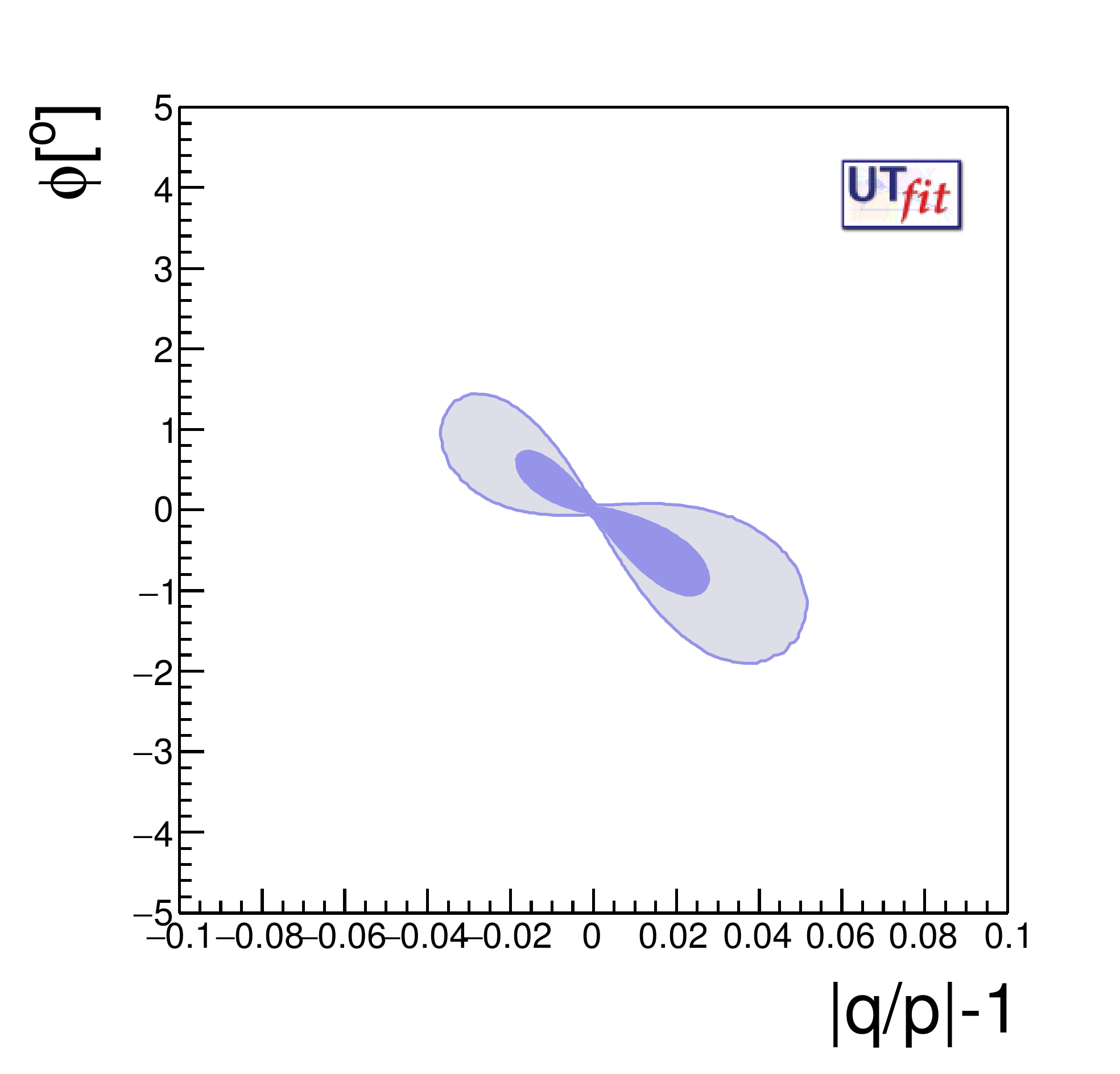} 
\includegraphics[width=0.3\textwidth]{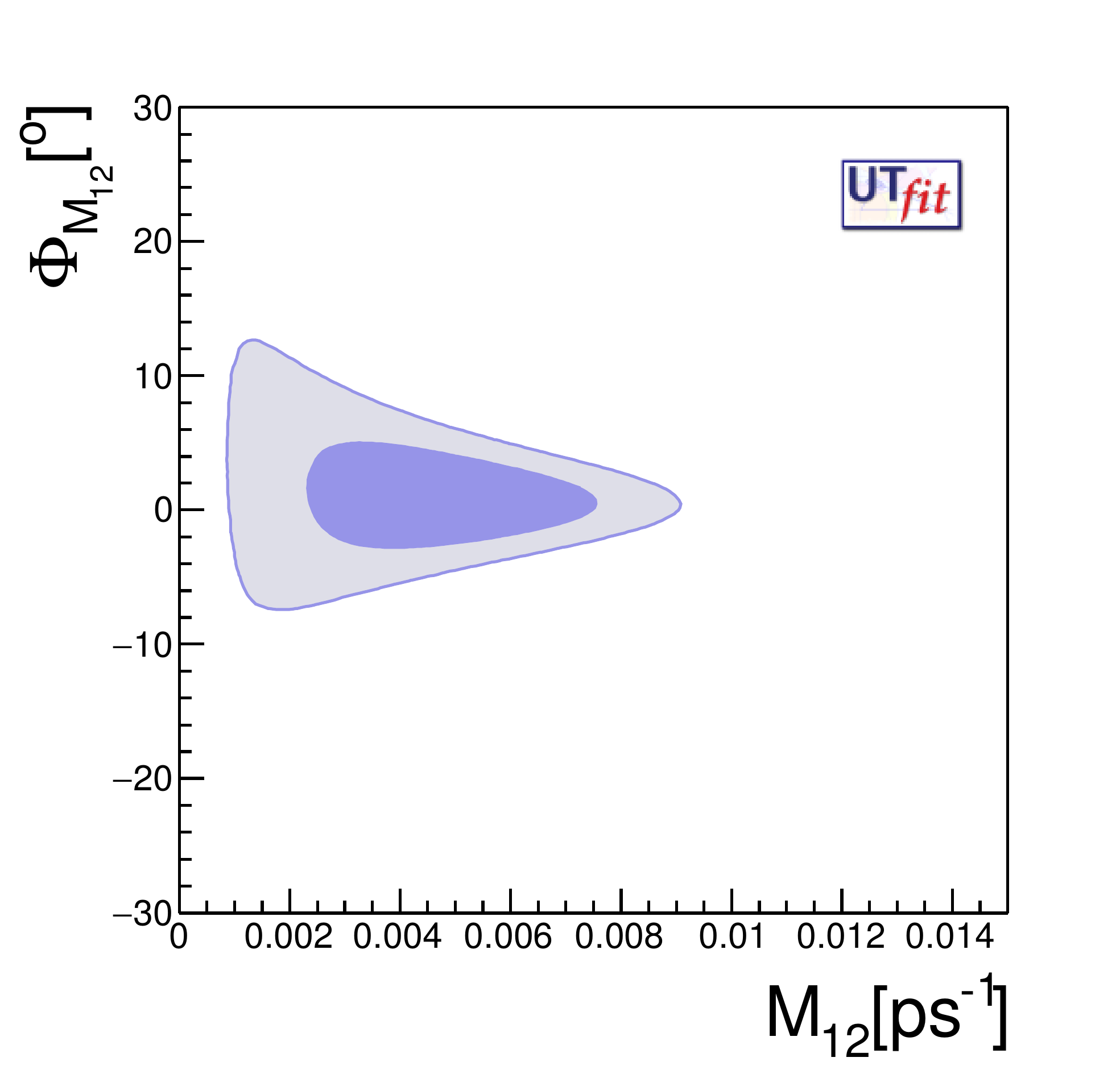} 
\caption{Results of the updated fit to $D$ mixing data. See
  ref.~\cite{Bevan:2014tha} for details.}
\label{fig:Dmix}
\end{figure}

The results of a global fit assuming real SM contributions and
searching for NP CP-violating effects by the UTfit Collaboration
\cite{Bevan:2014tha} are presented in Fig.~\ref{fig:Dmix} and in Table
\ref{tab:dmix}; see ref.~\cite{schwarz} for the updated HFAG fit. The
results show no evidence of CP violation within the current
experimental uncertainty, and allow to put severe bounds on the NP
scale. 

\begin{table}[htb]
\begin{center}
\begin{tabular}{l|ccc}  
    parameter & result @ $68\%$ prob. & $95\%$ prob. range\\
    \hline
    $\vert M_{12}\vert$ [ps$^{-1}$] & $(4.3 \pm 1.8) \cdot 10^{-3}$ & $[0.6,7.5]
    \cdot 10^{-3}$\\
    $\vert \Gamma_{12}\vert$ [ps$^{-1}$] & $(14.1 \pm 1.4) \cdot 10^{-3}$ &
    $[11.1,17.3] \cdot 10^{-3}$\\
    $\Phi_{M_{12}}$ [$^\circ$] & $(0.8 \pm 2.6)$ &
    $[-5.8,8.8]$\\
    \hline
    $x$ & $(3.5\pm 1.5) \cdot 10^{-3}$ & $[0.5,6.3]\cdot 10^{-3}$ \\
    $y$ & $(5.8\pm 0.6) \cdot 10^{-3}$ & $[4.5,7.1]\cdot 10^{-3}$ \\
    $\vert q/p\vert -1$ & $0.007\pm 0.018$ & $[-0.030,0.045]$ \\
    $\phi [^\circ]$ & $-0.21\pm 0.57$ & $[-1.53,1.02]$\\ \hline
\end{tabular}
\caption{Results of the fit to $D$ mixing data. See ref.~\cite{Bevan:2014tha} for details.}
\label{tab:dmix}
\end{center}
\end{table}

\begin{table}[htb]
\begin{center}
\begin{tabular}{l|ccc|}
 & $95\%$ upper limit  &
Lower limit on $\Lambda$ \\
&(GeV$^{-2}$) &
 (TeV)\\
\hline
$\mathrm{Im} C_D^1$  & $[-1.4,2.0] \cdot 10^{-14}$  &  $7.1\cdot 10^{3}$  \\ 
$\mathrm{Im} C_D^2$  & $[-2.5,1.7] \cdot 10^{-15}$  &  $20.0\cdot 10^{3}$ \\ 
$\mathrm{Im} C_D^3$  & $[-2.4,3.5] \cdot 10^{-14}$  &  $5.3\cdot 10^{3}$  \\ 
$\mathrm{Im} C_D^4$  & $[-5.2,7.7] \cdot 10^{-16}$  &  $36.0\cdot 10^{3}$  \\ 
$\mathrm{Im} C_D^5$  & $[-5.3,7.9] \cdot 10^{-15}$  &  $11.2\cdot 10^{3}$  \\ 
\hline
\end{tabular}
\end{center}
\caption{$95\%$ probability intervals for the imaginary part of the 
  Wilson coefficients, Im\,$C^D_i$,
  and the corresponding lower bounds on the NP scale, $\Lambda$, for
  a generic strongly interacting NP with generic flavor structure 
  ($L_i=F_i=1)$.} 
\label{tab:dmixnp}
\end{table}

The most general effective weak Hamiltonian for $D$ mixing of
dimension six operators is parameterized by Wilson coefficients of the
form
\begin{equation}
  C_i (\Lambda) = \frac{F_i L_i}{\Lambda^2}\, ,\qquad i=1,\ldots,5\, ,
  \label{eq:cgenstruct}
\end{equation}
where $F_i$ is the (generally complex) relevant NP flavor coupling,
$L_i$ is a (loop) factor which depends on the interactions that
generate $C_i(\Lambda)$, and $\Lambda$ is the NP scale, i.e.\ the
typical mass of new particles mediating $\Delta C=2$ transitions. For
a generic strongly interacting theory with an unconstrained flavor
structure, one expects $F_i \sim L_i \sim 1$, so that the
phenomenologically allowed range for each of the Wilson coefficients
can be immediately translated into a lower bound on
$\Lambda$. Specific assumptions on the flavor structure of NP
correspond to special choices of the $F_i$ functions. Assuming $F_i
=1$ and $L_i =
1$ and using the matrix elements recently computed in Lattice QCD
\cite{Carrasco:2014uya}, we obtain the bounds on the NP scale reported
in Table \ref{tab:dmixnp}. See ref.~\cite{Carrasco:2014uya} for details.

As anticipated above, the current uncertainty on  $\Phi_{12} = (0.8
\pm 2.6)^\circ$ is certainly compatible with the assumption of real SM
amplitudes. However, in view of the expected experimental progress,
it is mandatory to understand how one could go beyond this
assumption. As discussed in detail in ref.~\cite{kagan}, based on the
enhancement factor $1/\epsilon$ in eq.~(\ref{eq:uspin}) which is
absent in individual decay amplitudes, the dominant CP violating
effect in the SM can be captured by adding a universal phase
$\phi_{\Gamma_{12}}$ and fitting for both $\phi_{M_{12}}$ and
$\phi_{\Gamma_{12}}$. With present data we are not sensitive to
$\phi_{\Gamma_{12}}$ yet, but extrapolating to the expected
experimental accuracies after LHCb upgrade we foresee a determination
of $\phi_{\Gamma_{12}}$ with an error of $2^\circ$ and of
$\phi_{M_{12}}$ with an error of $1^\circ$. 

In addition to searching for CPV in $D$ mixing, another very clean
probe of NP is given by lepton number violating $D$ decays such as
$D^+_{(s)} \to \pi^- \mu^+ \mu^+$. These decay modes are very
sensitive to the presence of Majorana neutrinos with mass up to $1.1$
GeV, although for masses lower than around $400$ MeV Kaon decays
provide more stringent constraints. 

Last but not least among the clean probes of NP let me mention $D \to
\mu^+ \mu^-$. Within the SM, this decay is dominated by long-distance
contributions, but these can be reliably estimated once a measurement
of (or an upper bound on) BR$(D \to \gamma\gamma)$ is available. One
can then extract tight constraints on NP-induced short-distance $c
\bar{u} \to \mu^+ \mu^-$ transitions. See ref.~\cite{fajfer} for more
details. 

\section{NP-sensitive, theoretically challenging observables}
\label{sec:dirty}

Let us close this quick overview with a few potentially NP-sensitive
observables that however require substantial theoretical advance to
exploit their NP sensitivity. Generally speaking, all $\Delta C=1$
transitions with hadrons in the final state pose serious theoretical
challenges. The evaluation of (non-local) matrix elements is
problematic since charm is not heavy enough to apply QCD
factorization. Thus, waiting for lattice QCD to attack nonleptonic
charm decays, we can either look for possible order-of-magnitude NP
effects that could emerge over hadronic uncertainties, or try to
eliminate hadronic matrix elements using symmetry arguments.

For example, NP could give order-of-magnitude enhancements of the
long-distance dominated $D \to P \ell^+ \ell^-$ decays, leading to
bounds on NP contributions from the recent LHCb upper bounds on $D^+
\to \pi^+ \mu^+ \mu^-$ \cite{fajfer,deboer,Aaij:2013sua}. 

CP violation in Singly Cabibbo Suppressed (SCS) $D$ decays is
potentially sensitive to NP contributions, since SM contributions are
suppressed by the small CKM ratio $r = 6.5 \cdot 10^{-4}$. However, a
reliable estimate of the relevant hadronic matrix elements is needed
to identify possible NP contributions, unless one observes CP
asymmetries much larger than $10^{-3}$ or is able to get rid of the
unknown matrix elements using flavour symmetries. An interesting
example of the latter possibility is to study CP violation in
$\Delta I=3/2$ amplitudes, since no observable CPV is expected in
these amplitudes in the SM \cite{zupan}. 

The situation gets much more complicated if one wants to look for NP
in $\Delta I=1/2$ amplitudes. One could think of using
$\mathrm{SU}(3)$ to estimate the relevant matrix elements. However,
assuming exact SU$(3)$ symmetry one is not able to reproduce the
observed branching ratios \cite{paul,Hiller:2012xm}, and once SU$(3)$
breaking is allowed all possible reduced matrix elements are
generated, so no prediction is possible, except for a few sum rules
valid to second order in SU$(3)$ breaking
\cite{Grossman:2012ry,Gronau:2013xba,Gronau:2015rda}. Thus, while
SU$(3)$ can help identifying a hierarchy between the different
amplitudes, additional dynamical infomation is needed to predict CP
violation in SCS decays \cite{paul}. Several interesting attempts have
been made in this direction, using factorization
\cite{Buccella:1990sp,Buccella:1992ym,Buccella:1994nf,Wu:2004ht,Gao:2006nb},
dynamical assumptions about final state interactions
\cite{Buccella:2013tya} or $1/N_c$ arguments
\cite{schacht,Muller:2015lua,Muller:2015rna}. However, some degree of
model dependence is present in all these approaches, making it
difficult to reliably assess the uncertainty of the theoretical
predictions. Hopefully, with more experimental data and more
theoretical efforts this problem will be overcome in the near future.

\section{Conclusions}
\label{sec:concl}

I hope that this brief summary has stimulated the reader to look into
the details of the many interesting theoretical presentations that
have made CHARM-2015 such a lively conference. Charm physics is
playing a key role in improving our understanding of SM dynamics and
of what lies beyond the SM, and this role will be even more important
in the near future thanks to the foreseen experimental and theoretical
developments. Therefore I am sure that CHARM-2016 will be an even more
exciting conference.

\Acknowledgements It is a pleasure to thank the local organizing
committee for the very friendly and lively atmosphere at the
workshop. I am indebted to A. Lytle, V. Mateu and E. Gamiz for
granting permission to use their plots in this talk.

\bibliographystyle{JHEP}
\bibliography{charm2015_silvestrini}
 
\end{document}